\documentclass[pre,aps,twocolumn,showpacs]{revtex4}
\usepackage{psfrag,graphicx,epsfig,amsmath,amsfonts}

\def\vec#1{{\bf #1}}
\def\text#1{{\mathrm{#1}}}
\def\trace{{\mathrm{Tr}}}
\def\iff{{\rm ~if~~}}
\def\iff{{\rm ~~if~~}}
\def\ifff{{\rm ~if~~}}

\begin{document}

\title{The Kolmogorov-Sinai entropy for dilute systems of hard particles in equilibrium}
\author{Astrid S. de Wijn}
\email{A.S.deWijn@phys.uu.nl}
\affiliation{Institute for Theoretical Physics, Utrecht University, Leuvenlaan 4, 3584 CE, Utrecht, The Netherlands}

\begin{abstract}
\noindent
In an equilibrium system, the Kolmogorov-Sinai entropy, $h_{\mathrm{KS}}$, equals the sum of the positive Lyapunov exponents, the exponential rates
of divergence of infinitesimal perturbations.
Kinetic theory may be used to calculate the Kolmogorov-Sinai entropy for dilute gases of many hard disks or spheres in equilibrium at low number density $n$.
The density expansion of $h_{\mathrm{KS}}$ is $N \bar\nu A [\ln n + B + O(n)]$, where $\bar\nu$ is the single-particle collision frequency.
Previous calculations of $A$ were succesful.
Calculations of $B$, however, were unsatisfactory.
In this paper, I show how the probability distribution of the stretching factor can be determined from a nonlinear differential equation by an iterative method.
From this the Kolmogorov-Sinai entropy follows as the average of the logarithm of the stretching factor per unit time.
I calculate approximate values of $B$ and compare these to results from existing simulations.
The agreement is good.
\end{abstract}

\date{\today}

\maketitle

\section{Introduction}
Chaotic properties of systems with many degrees of freedom, such as moving hard spheres or disks, have been studied frequently.
Extensive simulation work has been carried out on their Lyapunov spectrum \cite{posch1,forster,christina}, and for low densities analytic calculations have been done for the largest Lyapunov exponent \cite{prlramses,
ramses,leiden,jstatph}, the Kolmogorov-Sinai entropy
\cite{prlramses,lagedichtheid}, and for the smallest positive Lyapunov exponents \cite{mareschal,onszelf}.

The Kolmogorov-Sinai entropy describes the maximal rate at which the system produces information about its phase-space trajectory.
In closed systems, it equals the sum of all positive Lyapunov exponents.

In this paper, I consider a system consisting of $N$ hard, spherical particles at small number density $n$, in $d$ dimensions ($d=2,3$).
I calculate the Kolmogorov-Sinai entropy in the low density approximation, where it is expected to behave as
\begin{align}
h_{\mathrm{KS}} = N \bar\nu A \left[\ln (n a^d) + B + O(na^d)\right]~,
\end{align}
where $\bar\nu$ is the average single-particle collision frequency, and $a$ is the particle diameter.
The constant $A$ has been calculated by van Beijeren et al.~\cite{lagedichtheid}, but the results found for $B$ were unsatisfactory.
I will show that the low-density approximation made in that calculation, which produces good results in the case of the Lorentz gas, which consists of uniformly convex scatterers, is too drastic in the case of a many-particle system.
This has already been anticipated in reference \cite{lagedichtheid} and preliminary estimates of the corrections were made by Dorfman in \cite{logtermen}, but the present calculations yield much more accurate values, which in principle could be improved even further.

The paper is organized as follows:
After a short introduction to Lyapunov exponents, the Kolmogorov-Sinai entropy will be introduced and its relation to the stretching factor will be discussed in Sec.~\ref{sec:lyap}, followed by an explanation of the relevant dynamics of hard disks in Sec.~\ref{sec:spheresdyn}.
In Sec.~\ref{sec:sf}, the stretching factor is calculated, and in Sec.~\ref{sec:dist} approximations are introduced to the probability distribution of the stretching factor.
The paper finishes with a discussion of the results in Sec.~\ref{sec:results}.

\section{\label{sec:lyap}Lyapunov exponents}
Consider a system with an $\cal N$-dimensional phase space $\Gamma$.
At time $t=0$, the system is at an initial point ${\vec\gamma}_0$ in this space.
It evolves with time according to ${\vec\gamma}({\vec\gamma}_0,t)$.
If the initial conditions are perturbed infinitesimally, by $\delta{\vec\gamma}_0$, the system evolves along an infinitesimally different path $\gamma + 
\delta \gamma$, which can be specified by
\begin{align}
{\delta{\vec\gamma}({\vec\gamma}_0,t)} \label{eq:M}= {{\sf M}_{{\vec\gamma}_0}(t)\cdot \delta{\vec\gamma}_0~,}
\end{align}
in which the matrix ${\sf M}_{{\vec\gamma}_0}(t)$ is defined by
\begin{align}
\label{eq:tang} {\sf M}_{{\vec\gamma}_0}(t)=\frac{d {\vec\gamma}({\vec\gamma}_0,t)}{d {\vec\gamma}_0}~.
\end{align}
The Lyapunov exponents are the possible average rates of growth of such perturbations, i.e.,
\begin{equation}
\lambda_i = \lim_{t\rightarrow\infty} \frac{1}{t} \ln 
|\mu_i(t)|~,
\end{equation}
where $\mu_i(t)$ is the $i$-th eigenvalue of ${\sf M}_{{\vec\gamma}_0}(t)$.
If the system is ergodic, it comes arbitrarily close to any point in phase space for all initial conditions except for a set of measure zero.
Therefore, the Lyapunov exponents are the same for almost all initial conditions.
One may order the exponents according to size, with $\lambda_1$ being the largest and $\lambda_{\cal N}$ the smallest, as is the convention.
For each exponent there is a corresponding eigenvector of ${\sf M}_{{\vec\gamma}_0}(t)$.

For a classical system of $N$ $d$-dimensional freely moving hard spheres without internal degrees of freedom, the phase space and tangent space may be represented by the positions and velocities of all particles and their infinitesimal deviations,
\begin{align}
\gamma_i & = (\vec{r}_i, \vec{v}_i)~,\\
\delta\gamma_i & =  ({\vec{ \delta r}}_i, {\vec{ \delta v}}_i)~,
\end{align}
where $i$ runs over all particles and $\delta\gamma_i$ is the contribution of particle $i$ to $\delta\gamma$.

\subsection{\label{sec:ks}Kolmogorov-Sinai entropy and stretching factor}

In standard terminology, the stretching factor $\Lambda(t)$ is defined as the factor by which the expanding part of tangent space stretches over a time $t$.
This quantity can be used to calculate the Ruelle pressure as well as the sum of the positive Lyapunov exponents, which is equal to the Kolmogorov-Sinai entropy in systems without escape \cite{henkenbob1,henkenbob2,onslorentz}.
For long times, the stretching is dominated by the positive Lyapunov exponents, and one has for the Kolmogorov-Sinai entropy
\begin{align}
h_{\mathrm{KS}} =  \lim_{t\rightarrow \infty} \frac{1}{t} \ln \Lambda(t)~.
\label{eq:ks}
\end{align}
For long times, the stretching factor can be calculated from the total growth of an arbitrary volume element in $dN$ dimensions.
After a few multiples of the inverse of the smallest positive Lyapunov exponent, the dynamics project the volume element onto the expanding manifold and its subsequent growth is completely described by the stretching factor.

For hard-sphere systems, where the collision times are defined exactly, the stretching factor can be written as the product of the stretching factors resulting from each of the different single collisions combined with the subsequent (or previous) free flights of the two particles involved.
In this description, the effects of the free flights of the other particles are accounted for at the collisions involving those particles.
On the right-hand side of Eq.~(\ref{eq:ks}), the logarithm may be replaced by the sum of logarithms of these single-collision stretching factors.
The resulting expression may be interpreted as a time average, which in ergodic systems may be replaced by an ensemble average.
Hence,
\begin{align}
h_{\mathrm{KS}} =  \frac{N\bar\nu}{2}\langle \ln \Lambda_i\rangle~.
\label{eq:gemiddeld}
\end{align}
At low densities, the single-particle collision frequency, $\bar\nu$, is given by
\begin{align}
\bar\nu = \frac{2 \pi^{(d-1)/2}}{\Gamma\left(\frac{d}{2}\right)}\frac{n a^{d-1}}{\sqrt{\beta m}}~.
\label{eq:nu}
\end{align}
The factor $N\bar\nu/2$ in Eq.~(\ref{eq:gemiddeld}) equals to the overall collision frequency.
$\Lambda_i$ is the single-collision stretching factor due to collision $i$.
In this paper, this includes the effects of the free flights of the colliding particles after the collision, and not those before.
To obtain the Kolmogorov-Sinai entropy from this, one must calculate the distribution of single-collision stretching factors.

\section{\label{sec:spheresdyn}Dynamics of hard spheres in phase space and tangent space}
\begin{figure}
\includegraphics[width=6cm]{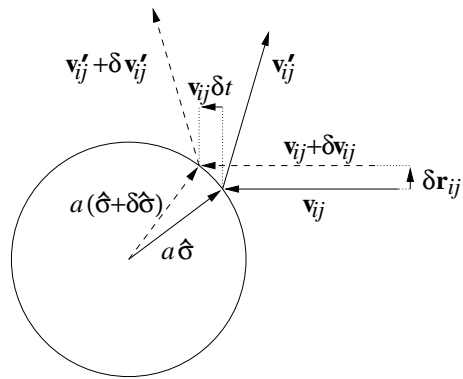}
\caption{\label{fig:bolletje} Geometry of a collision of two particles in relative coordinates.  The collision normal $\hat{\vec{\sigma}}$ is the unit vector pointing from the center of one particle to the center of the other.}
\end{figure}

In order to calculate the single-particle stretching factors, one must derive the dynamics of the system in tangent space from the dynamics in phase space.

The evolution in phase space consists of an alternating sequence of free flights and collisions.
During free flights the particles do not interact and the positions grow linearly with the velocities,
\begin{align}
\label{eq:freeflight1}\vec{r}_i(t) &= \vec{r}_i(t_0) + (t - t_0)\vec{v}_i(t_0)~,\\
\vec{v}_i(t) &= \vec{v}_i(t_0)~.
\end{align}
At a collision, momentum is exchanged between the colliding particles along the collision normal, $\hat{\vec{\sigma}} = (\vec{r}_i-\vec{r}_j)/{a}$, as shown in Fig.~{\ref{fig:bolletje}.
The other particles do not interact.  Using primes to denote the coordinates in phase space after the collision, one finds
\begin{align}
\vec{r}'_i &= \vec{r}_i~,\\
\label{eq:deltav}
\vec{v}'_i &= \vec{v}_i - \hat{\vec{\sigma}} (\hat{\vec{\sigma}} \cdot \vec{v}_{ij})~,
\end{align}
where $\vec{v}_{ij}=\vec{v}_i-\vec{v}_j$.
From Eq.~(\ref{eq:tang}) and Eqs.~(\ref{eq:freeflight1})-(\ref{eq:deltav}) the dynamics in tangent space can be derived \cite{ramses}.

During free flight there is no interaction between the particles and the components of the tangent-space vector transform according to
\begin{align}
\left(
\begin{array}{c}
\delta\vec{r}_i'\\
\delta\vec{v}_i'
\end{array}
\right)
&= 
\left(\begin{array}{cc}
\extracolsep{1mm}
\rule{0mm}{0mm}{\bf 1}&(t-t_0){\bf 1}\\
\rule{0mm}{5mm}0&{\bf 1}
\end{array}\right)
\cdot \left(
\begin{array}{c}
\delta\vec{r}_{i}\\
\delta\vec{v}_{i}
\end{array}
\right)
~,
\label{eq:flight}
\end{align}
in which ${\bf 1}$ is the $d\times d$ identity matrix.

A collision between particles $i$ and $j$ only changes the tangent-space vectors of the colliding particles \cite{prlramses}.
As shown in Fig.~\ref{fig:bolletje}, an infinitesimal difference in the positions of the particles leads to an infinitesimal change in the collision normal and in the collision time.
The $\vec{v} + \delta \vec{v}$ are exchanged along $\hat{\vec\sigma}+\delta\hat{\sigma}$ according to Eq.~(\ref{eq:deltav}).
This leads to infinitesimal changes in both positions and velocities right after the collision.
For convenience one may switch to relative and center of mass coordinates, $\delta\vec{r}_{ij} = \delta\vec{r}_i -\delta\vec{r}_j, \delta\vec{R}_{ij} = (\delta\vec{r}_i +\delta\vec{r}_j)/2, \delta\vec{v}_{ij} = \delta\vec{v}_i -\delta\vec{v}_j$, and
$\delta\vec{V}_{ij} = (\delta\vec{v}_i +\delta\vec{v}_j)/2$.
These transform as
\begin{align}
\label{eq:reldyn1}
\delta\vec{r}'_{ij} &= \delta\vec{r}_{ij} - 2 {\sf S} \cdot \delta\vec{r}_{ij}~,\\
\delta\vec{R}'_{ij} &= \delta\vec{R}_{ij}~,\\
\delta\vec{v}'_{ij} &= \delta\vec{v}_{ij} - 2 {\sf S} \cdot \delta\vec{v}_{ij} - 2 {\sf Q} \cdot \delta\vec{r}_{ij}~,\\
\delta\vec{V}'_{ij} &= \delta\vec{V}_{ij}~,
\label{eq:reldyn4}
\end{align}
in which ${\sf S}$ and ${\sf Q}$ are the $d \times d$ matrices
\begin{align}
\label{eq:S}{\sf S} & =  \hat{\vec{\sigma}} \hat{\vec{\sigma}}~,\\
\label{eq:Q}{\sf Q} & =  \frac{\left[(\hat{\vec{\sigma}}\cdot\vec{v}_{ij})\,{\bf 1}+\hat{\vec{\sigma}}\vec{v}_{ij}\right]\cdot
                     \left[ (\hat{\vec{\sigma}}\cdot\vec{v}_{ij})\,{\bf 1}-\vec{v}_{ij} \hat{\vec{\sigma}}  \right]}
               {a (\hat{\vec{\sigma}}\cdot\vec{v}_{ij})}~.
\end{align}
Here the notation $\vec{a}\vec{b}$ denotes the standard tensor product of vectors $\vec{a}$ and $\vec{b}$.
Note that ${\sf Q}$ transforms $\delta\vec{r}_{ij}$ vectors which are orthogonal to $\vec{v}_{ij}$ into vectors which are orthogonal to $\vec{v}'_{ij}$.
The vector $\hat{\vec{v}}_{ij}$ is a right zero eigenvector of ${\sf Q}$, and $\hat{\vec{v}}'_{ij}$ a left zero eigenvector.
Note that these are $d$-dimensional vectors, not $2d$-dimensional.

\section{\label{sec:sf}Stretching factor}

In order calculate the Kolmogorov-Sinai entropy from Eq.~(\ref{eq:gemiddeld}), one must find the probability distribution of the single-particle stretching factor.
In this section, I first derive the single-collision stretching factor as a function of the collision parameters and other parameters which contain information about the history of the system.
Information about the history will be replaced by a pre-collisional distribution function, which is the distribution function averaged over an ensemble of initial conditions.

\subsection{Projection}

The growth of a $dN$-dimensional volume element in $\delta\Gamma$ can be monitored through its projection onto a subspace of $\delta\Gamma$ with at least the same number of dimensions, as long as this projection space is not orthogonal to one of the $dN$ leading eigenvectors of ${\sf M}$.
In the limit $t\rightarrow \infty$, the logarithm of the determinant of the transformation of the projection yields the same Kolmogorov-Sinai entropy as the logarithm of the stretching factor of the actual volume element.

If {$(\delta\vec{r}_i^{(m)},\delta\vec{v}_i^{(m)})$} are the eigenvectors belonging to the positive exponents, the eigenvectors which belong to their counterparts under conjugate pairing are equal to {$(\delta\vec{r}_i^{(m)},-\delta\vec{v}_i^{(m)})$}.
This means that eigenvectors which have no contributions along either $\delta\vec{r}_i$ or $\delta\vec{v}_i$ correspond to themselves under conjugate pairing.
Such eigenvectors must therefore have Lyapunov exponents which are zero.
The $dN$-dimensional vectors whose components belonging to particle $i$ are $\delta\vec{r}_i$ respectively $\delta\vec{v}_i$ are denoted by
The spaces spanned by either $\delta\vec{r}$ or $\delta\vec{v}$ are not orthogonal to any eigenvectors which belong to nonzero Lyapunov exponents.
$\delta\vec{r}$ and $\delta\vec{v}$.
In the system described here a convenient choice for the projection space may therefore be either of these spaces.
Here, $\delta\vec{v}$ is used for this purpose, because it does not change during free flights.

\subsection{Stretching factor of a single collision}

During a free flight, $\delta\vec{r}$ grows with $\delta\vec{v}$.
Denoting the perturbations in the position just after a collision with a superscript $+$ and those just before the next collision with a superscript $-$, one may write
\begin{align}
\delta\vec{r}_i^- = \delta\vec{r}_i^+ + \tau_i \delta\vec{v}_i~,
\end{align}
where $\tau_i$ is the free-flight time of particle $i$.
Note that $\tau_i$ typically is of the order of $1/\bar\nu$.
In previous calculations, it was usually assumed that right after a collision $\delta\vec{r}^+_i$ and $\delta\vec{v}_i$ were of the same order of magnitude \cite{lagedichtheid}.
Under this assumption, the contribution from $\delta\vec{r}^+_i$ to $\delta\vec{r}^-_i$ may be neglected compared to $\tau_i \delta\vec{v}_i$.
Of course $\tau_i\delta\vec{v}_i$ will be comparable to $\delta\vec{r}^+_i$ after the previous collision if $\tau_i$ is short.
However, this occurs only with a probability proportional to the density and therefore may be neglected in the average.

The assumption that $\delta\vec{r}_i$ and $\delta\vec{v}_i$ just after a collision are of the same order of magnitude, however, is only true for $d-1$ components of $\delta\vec{r}_{ij}$, namely the ones normal to $\hat{\vec{v}}_{ij}$.
The remaining component of $\delta\vec{r}_{ij}$, which is along $\hat{\vec{v}}_{ij}$, and all components of $\delta\vec{R}_{ij}$ are, after a collision, larger by an order of $\tau$ than the corresponding component of $\delta\vec{v}_i$, because ${\sf Q}$, defined in Eq.~(\ref{eq:Q}), does not act on centre-of-mass perturbations, nor on perturbations of relative coordinates parallel to the relative velocity.
In these directions, the components of $\delta\vec{v}$ are of the same order of magnitude as before the collision, but the corresponding components of $\delta\vec{r}$ have grown linearly during the preceding free flights.
I will show that this affects the Kolmogorov-Sinai entropy, even at low density.

The determinant of the transformation of the $dN$-dimensional volume element projected onto $\delta\vec{v}$ depends on $\delta\vec{r}$ before the collision.
$\delta\vec{r}$ may be assumed to depend on $\delta\vec{v}$ as
\begin{align}
\delta\vec{r} =  \bar\tau {\mathcal W}\cdot \delta\vec{v}~,
\label{eq:mathcalW}
\end{align}
 with $\bar\tau = 1/\bar\nu$ the average free-flight time.
${\mathcal W}$ is proportional to the inverse of the radius of curvature tensor, which is often used to calculate Lyapunov exponents (see, for example, \cite{henkenbob1}).
The matrix ${\mathcal W}$ can be split up into $d\times d$ matrices between specific particles, ${\sf W}_{ij}$.
As particles collide and have free flights, ${\sf W}_{ij}$ changes.
The volume element projected onto $\delta\vec{v}$ before the collision is mapped to a projection of a volume element after the collision.
The determinant of this map depends on ${\sf W}_{i}, {\sf W}_{j}, {\sf W}_{ij}$, and ${\sf W}_{ji}$, where the second index is omitted if it is the same as the first.

After the collision, ${\mathcal W}$ is changed.
From now, a prime will be used to denote a quantity after a collision and the subsequent free flight(s).
Quantities just after a collision, before the subsequent free flight are indicated with a superscript $*$.

The matrix ${\mathcal W}$ after the collision, ${\mathcal W}^*$, can be found by using the dynamics and Eq.~(\ref{eq:mathcalW}) to express $\delta\vec{r}^*$ just after the collision in terms of $\delta\vec{v}^*$, the collision matrices and ${\mathcal W}$.
Let ${\mathcal S}$ and ${\mathcal Q}$ be the $dN \times dN$-dimensional matrices which perform the transformations of ${\sf S}$ and ${\sf Q}$ on the components of $\delta\vec{r}$ and $\delta\vec{v}$ along the colliding particles, as described in Eqs.~(\ref{eq:reldyn1})--(\ref{eq:reldyn4}) and act as the unit operator on the components belonging to all other particles.
The transformation can be written as
\begin{align}
\delta\vec{v}' &= ({\mathcal I}+{\mathcal S})\cdot\delta\vec{v} + {\mathcal Q}\cdot\delta\vec{r}\\
\label{eq:transdvdv}
& = ({\mathcal I}+{\mathcal S}+\bar\tau{\mathcal Q}\cdot{\mathcal W})\cdot\delta\vec{v} ~, \\[1ex]
\delta\vec{r}^*
 &= ({\mathcal I}+{\mathcal S})\cdot\delta\vec{r}
 = \bar\tau({\mathcal I}+{\mathcal S})\cdot {\mathcal W}\cdot \delta\vec{v}~.
\end{align}
Here, ${\mathcal I}$ is the $dN\times dN$ identity matrix.
Note that $\delta\vec{v}'$ is equal to $\delta\vec{v}^*$, and that all components of $\delta\vec{r}^*$ and $\delta\vec{r}'$ that do not belong to either of the colliding particles remain the same.
This leads to an expression for $\delta\vec{r}^*$ as a function of $\delta\vec{v}^*$.
\begin{align}
\delta\vec{r}^* = \bar\tau{\mathcal W}^* \cdot \delta\vec{v}^*~,
\end{align}
where ${\mathcal W}^*$ can be expressed in terms of ${\mathcal W}$ and the collision matrices, as
\begin{align}
{\mathcal W}^* = ({\mathcal I}+{\mathcal S})\cdot {\mathcal W}\cdot ({\mathcal I}+{\mathcal S}+\bar\tau{\mathcal Q}\cdot{\mathcal W})^{-1}~.
\end{align}
Using $({\mathcal I}+{\mathcal S})^{-1}= {\mathcal I}+{\mathcal S}$, one may write this more conveniently as
\begin{align}
{\mathcal W}^* 
&= ({\mathcal I}+{\mathcal S})  \cdot [{\mathcal W}^{-1} + \bar\tau ({\mathcal I}+{\mathcal S}) \cdot {\mathcal Q}]^{-1} \cdot ({\mathcal I}+{\mathcal S})~.
\label{eq:mathcalWa}
\end{align}

At low densities, two particles which collide can be assumed to be uncorrelated before the collision (Sto{\ss}zahlansatz).
This means that ${\sf W}_{ij}= {\sf W}_{ji}=0$, if $i\neq j$.
After the collision there generally are nonzero elements in ${\sf W}^*_{ij}$.

${\sf W}_i$ also changes during the subsequent free flight.
Let $\tau_i$ be the free-flight time of particle $i$ after the collision.  Then, after the free flight,
\begin{align}
{\sf W}'_k = 
\begin{cases}
{\sf W}^*_k + {\bf 1} \bar\nu\tau_k& \iff k=i,j~,\\
{\sf W}^*_k& \iff k\neq i,j~.
\end{cases}
\label{eq:mathcalWt}
\end{align}
Note that the change in $\delta\vec{r}_k$ with $k\neq i,j$ due to free flights was already taken into account at the collisions of particle $k$.
The matrix ${\mathcal W}$ as it is calculated here describes the connections between $\delta\vec{r}_i$ just before the next collision of particle $i$ and $\delta\vec{v}_j$ just before the next collision of particle $j$.

The matrix $({\mathcal I}+{\mathcal S}) \cdot  {\mathcal Q}$ is nonnegative definite and symmetric.
If ${\mathcal W}$ is positive definite, so is its inverse.
This means that ${\mathcal W}^{-1} + \bar\tau ({\mathcal I}+{\mathcal S}) \cdot {\mathcal Q}$ is positive definite, as is its inverse.
The coordinate reflection ${\mathcal I}+{\mathcal S}$ is unitary, and therefore the eigenvalues of $({\mathcal I}+{\mathcal S})\cdot {\mathcal W}^* \cdot ({\mathcal I}+{\mathcal S})$ are the same as those of ${\mathcal W}^*$.
Therefore, ${\mathcal W}^*$ is positive definite.
By similar reasoning, a symmetric ${\mathcal W}$ is mapped onto a symmetric ${\mathcal W}^*$.
Equation~(\ref{eq:mathcalWt}) also maps nonnegative definite matrices onto positive definite matrices, and symmetric ones onto symmetric ones.
As, without loss of generality, any initial conditions for ${\mathcal W}$ may be chosen, it is possible to choose them such that ${\mathcal W}$ is positive definite and symmetric.
This can be done, for example, by choosing the initial ${\mathcal W}$ to be diagonal, with elements equal to $1$.

The stretching factor due to one collision can be calculated from the determinant of the transformation of the projection onto the perturbations of the relative and centre-of-mass velocities, $\delta\vec{v}_{ij}$ and $\delta\vec{V}_{ij}$.
From Eq.~(\ref{eq:transdvdv}), one finds that this is the determinant of ${\mathcal I}+\bar\tau({\mathcal I} +{\mathcal S})\cdot{\mathcal Q}\cdot{\mathcal W}$.

In the low density limit, and with ${\sf W}_{ij}=0$, this is found to be equal to
\begin{align}
\Lambda = w_{\perp\perp}
\left(\frac{2 v \bar\tau}{a}\right)^{d-1} \cos^{d-3}\theta~.
\label{eq:sf}
\end{align}
Here, $\theta$ is the angle between $\hat{\sigma}$ and $\vec{v}$, $\cos\theta=\hat\sigma\cdot\hat{\vec{v}}$, and $w_{\perp\perp}$ is 
equal to the determinant of the part of ${({\sf W}_i + {\sf W}_j)/2}$ between vectors that are orthogonal to $\hat{\vec{v}}_{ij}$ before the collision.
For $d=2$,
\begin{align}
\label{eq:aperpperp}
w_{\perp\perp} = \hat{\vec{v}}_{ij \perp}\cdot\frac{{\sf W}_i + {\sf W}_j}{2}\cdot \hat{\vec{v}}_{ij \perp}~.
\end{align}
This expression replaces the factor $\bar\nu\tau_+$, where $\tau_+= (\tau_i+\tau_j)/2$, in previous calculations of the single-particle stretching factor \cite{lagedichtheid}.
In $d$ dimensions, $\hat{\vec{v}}_{ij \perp}$ must be replaced by a set of $d-1$ vectors orthogonal to $\hat{\vec{v}}_{ij}$.
The inner products are then replaced by the determinant of the $(d-1)\times(d-1)$ matrix with elements given by the innerproducts of $({\sf W}_i + {\sf W}_j)/2$ between those vectors.

\section{\label{sec:dist}Distribution functions}

In order to calculate the Kolmogorov-Sinai entropy from Eq.~(\ref{eq:gemiddeld}), one needs the distribution function of the single-collision stretching factor, as described by Eqs.~(\ref{eq:sf}) and (\ref{eq:aperpperp}).
This may be derived from the joint distribution function of the collision parameters $\tau_i$, $\tau_j$, $\vec{v}_i$, $\vec{v}_j$, $\theta$, and the elements of ${\sf W}_i$ and ${\sf W}_j$.
In the low-density approximation, the collision parameters are distributed according to the equilibrium solutions of the Boltzmann equation.

The distribution of the particle velocities is the Maxwell distribution,
\begin{align}
\phi_{\text{M}}(\vec{v}_i) = \left(\frac{ 2 \pi}{m \beta}\right)^{-d/2} \exp{\left( - \frac12 {\beta m |\vec{v}|^2}\right)}~.
\end{align}
The collision rate for collisions at angle $\theta$ and with outgoing velocities $\vec{v}_i$ and $\vec{v}_j$ is proportional to the differential cross section times the relative velocity, $\sin^{d-2}\theta \cos\theta v_{ij}$.
The normalized probability distribution of the collision parameters is thus equal to
\begin{align}
p(\vec{v}_i,\vec{v}_j,\theta)\,d\vec{v}_i d\vec{v}_j d\theta = &\sin^{d-2}\theta \cos\theta  \frac{|\vec{v}_i-\vec{v}_j|n a^{d-1}}{\bar\nu}\nonumber\\
&\times \phi_{\text{M}}(\vec{v}_i) \phi_{\text{M}}(\vec{v}_j)\,d\vec{v}_i d\vec{v}_j d\theta~.
\label{eq:distrelvel}
\end{align}
The free-flight times of the particles are distributed exponentially, with the collision frequency $\nu(\vec{v})$ depending on the velocity of the particle, according to
\begin{align}
p_\tau (\tau|v_i) d\tau = \nu({\vec{v}_i}) \exp[{-\nu({\vec{v}_i}) \tau}] d\tau~,
\end{align}
where $\nu(\vec{v}_i)$ is the velocity dependent collision frequency.
This can can be found by integrating the collision rate over the outgoing velocity of the other particle and over the collision normal, i.~e.
\begin{align}
\nu(\vec{v}_i) = \int d\vec{v}_j d\theta\, n a^{d-1} \sin^{d-2} \cos\theta|\vec{v}_i -\vec{v}_j| \phi_{\text{M}}(\vec{v}_j)~.
\end{align}

The distribution of $w_{\perp\perp}$ can be found from the requirement that the distribution of elements of ${\sf W}_{i}$ as a function of $\vec{v}_i$ is not changed by collisions.
This yields a complicated nonlinear differential equation for the distribution of the elements of ${\sf W}_i$ and ${\sf W}_j$.
It involves the distribution of angles between the relative velocities of subsequent collisions of a particle with velocity $\vec{v}_i$ as well as the velocity dependence of the collision frequency.
The latter is only known numerically.
With the implied inclusion of the collision parameters and coordinates of both particles in the distribution functions implied in the integral,
the equation can be written as
\begin{align}
p'(\tilde{\mathcal W}) = \int d{\mathcal W}\, p({\mathcal W}) \delta({\mathcal W}'({\mathcal W}) - \tilde{\mathcal W})~.
\label{eq:nonlin}
\end{align}
The solution to this equation can be approximated using an iterative approach.

\subsection{\label{sec:mean-fieldw}Approximation of the distribution of $w_{\perp\perp}$}

Rather than solving Eq.~(\ref{eq:nonlin}) exactly, which is not feasable, Eq.~(\ref{eq:nonlin}) may be used to iterate the distribution.
One may start with an initial distribution, $p({\mathcal W})$, and calculate the distribution after one collision, $p'({\mathcal W})$.
After every iteration, the distribution more closely resembles the true solution of the integral equation.
However, even with a simple initial distribution, such iterations will quickly produce distribution functions which can only be calculated numerically.
In principle, the equation could be such that there is no convergence at all, but the nature of the physical problem dictates that the distribution of ${\mathcal W}$ converges after many collisions.

In this section, an alternative iterative approach is used to find an approximate distribution function.
We start with a simple distribution with one parameter, which approximates the average trace element.
The parameter is chosen in such a way that the average of the trace of ${\mathcal W}$ remains the same after a collision and free flights.
The nonlinear terms in the equation for the distribution of the elements of ${\mathcal W}$, which have zero average, can be ignored at first, due to the choice of initial distribution.
The integral equation is iterated a second time to include some of these terms.
Subsequently, the size of the remaining corrections after more iterations is estimated.

\subsubsection{The trace of ${\mathcal W}$}

In principle, it would be possible to use the determinant or some other scalar function of ${\mathcal W}$, instead of the trace.
It is however much easier to write down the map of the trace of ${\mathcal W}$ onto the trace of ${\mathcal W}^*$ in Eq.~(\ref{eq:mathcalWa}) than it is to write down a map of the determinant during free flights.
Also, only the distribution of diagonal elements of ${\mathcal W}$ is actually needed.
Under unitary coordinate transformations, such as rotations and reflections, the trace of a matrix is conserved.
Using a parameter is not really necessary, however, it greatly improves the convergence toward the solution of the equation.

From Eq.~(\ref{eq:mathcalWt}), the trace of ${\mathcal W}'$ can be found to satisfy
\begin{align}
\trace({\mathcal W'}) &= \trace({\mathcal W}^*) + d \bar\nu\tau_i + d \bar\nu\tau_j~.
\label{eq:tracet}
\end{align}
Let the $dN$-dimensional basis vectors in which the matrices are expressed be numbered 1 through $dN$.
Let the first $dN$-dimensional basis vector, $\vec{\epsilon}_1$, be defined as $\hat{\vec{v}}_{ij}$ in the relative coordinates, and the second (in $d$ dimensions the second through $d$-th), $\vec{\epsilon}_2$, as $\hat{\vec{v}}_{ij\perp}$ in the relative coordinates.
The remaining basis vectors may be chosen in any arbitraty way, as long as they are orthogonal to eachother and of unit length.

Let ${\mathcal A}_{\ell}$ represent the $dN \times dN$ matrix ${\mathcal A}$ with all rows and columns removed except for those with indices specified by the list $\ell$, where $\ell$ may be any list of indices.
Similarly, let ${\mathcal A}_{(\ell)}$ be the matrix ${\mathcal A}$ with all rows and columns removed with indices belonging to the list $\ell$.
Specifically, ${\mathcal A}_{(k)}$ and ${\mathcal A}_{(k2)}$ represent the matrix ${\mathcal A}$ from which the rows and columns belonging to, respectively, index $k$ and both index $k$ and index $2$ are removed.

From Eq.~(\ref{eq:mathcalWa}), the trace of ${\mathcal W}^*$ can be found as a function of the elements of ${\mathcal W}$.
As the trace is conserved under the coordinate reflection $({\mathcal I} +{\mathcal S})\cdot{\mathcal W}\cdot ({\mathcal I} +{\mathcal S})$, one finds
\begin{align}
\trace({\mathcal W}^*) & = \sum_k \frac{\det ({\mathcal W}^{-1} + \bar\tau ({\mathcal I }+ {\mathcal S})\cdot {\mathcal Q})_{(k)}}{\det({\mathcal W}^{-1} + \bar\tau ({\mathcal I }+ {\mathcal S}) \cdot {\mathcal Q})}~.
\label{eq:trwster}
\end{align}
Here, the sum is over an orthonormal basis of $dN$ unit vectors.
In the low-density limit the mean free time becomes large, and only terms in which the numerator contains the same power of $\bar\tau$ as the denominator can contribute.
The product of the nonzero eigenvalues of $({\mathcal I} + {\mathcal S})\cdot {\mathcal Q}$ can be divided out, leaving only the determinant of the remaining part of ${\mathcal W}$, between vectors on which ${\mathcal Q}$ does not work.
As only $({\mathcal I}+ {\mathcal S}) \cdot {\mathcal Q}$ in Eq.~(\ref{eq:trwster}) contains the collision normal, the trace of ${\mathcal W}^*$ does, in the limit of vanashing density, not depend on $\theta$, but only on $\hat{\vec{v}}_{ij}$ and the elements of ${\mathcal W}$.

The trace of ${\mathcal W}^*$ can be rewritten as the sum over fractions of subdeterminants,
\begin{align}
\trace({\mathcal W}^*) & =  \sum_{k \neq 2}
\frac{\det({\mathcal W}^{-1}_{(k 2 )}) }
     {\det({\mathcal W}^{-1}_{  (2) }) }~.
\label{eq:sumtrace}
\end{align}
Further, after writing the inverse of ${\mathcal W}^{-1}_{\ell}$ and by working out the determinant of ${\mathcal W}_{\ell}$, by induction over the number of indices occurring in $\ell$ 
\begin{align}
\det {\mathcal W}^{-1}_{(\ell)} = \det {\mathcal W}_\ell \, \det {\mathcal W}^{-1}~.
\label{eq:detfrac}
\end{align}
From Eqs.~(\ref{eq:detfrac}) and (\ref{eq:sumtrace}) one finds, for $d=2$, 
\begin{align}
\trace({\mathcal W}^*) &= \sum_{k \neq 2} \frac{\det({{\mathcal W}_{k 2 } }) }{\det{\mathcal W}_{2}}  \\
&=  \trace({\mathcal W})  - \frac{1}{w_{\perp\perp}} \,{\vec{\epsilon}_2} \cdot {\mathcal W}^2\cdot {\vec{\epsilon}}_2~.
\label{eq:trace}
\end{align}
In the $d$-dimensional case, ${\vec{\epsilon}}_2$ is replaced by $d-1$ vectors, and Eq.~(\ref{eq:trace}) becomes somewhat more complicated.

The change in off-diagonal elements at a collision can be found from a derivation similar to that for the trace in Eq.~(\ref{eq:trace}),
\begin{align}
\vec{\epsilon}_p^* \cdot {\mathcal W}^* \vec{\epsilon}_q^* = \vec{\epsilon}_p \cdot {\mathcal W} \cdot \vec{\epsilon}_q - \frac{(\vec{\epsilon}_p \cdot {\mathcal W} \cdot \vec{\epsilon}_2)(\vec{\epsilon}_2 \cdot {\mathcal W} \cdot \vec{\epsilon}_q)}{\vec{\epsilon}_2 \cdot {\mathcal W} \cdot \vec{\epsilon}_2}~.
\label{eq:offdiagel}
\end{align}
The expression for $d=3$ is similar.
If $p$ or $q$ is equal to $2$, the off-diagonal element vanishes.
Off-diagonal elements between different particles are not affected by free flights, as is apparent from Eq.~(\ref{eq:mathcalWt}).

The collisions are most conveniently expressed in the basis which consists of $\hat{\vec{v}}_{ij}$ and the $d-1$ vectors $\hat{\vec{v}}_{ij\perp}$ orthogonal to it.
I therefore also express each ${\sf W}_{ij}$ in this basis.

\subsubsection{Iterative approach}

Assume that just before a collision the ${\sf W}_{i}$ are equal to their averages and ${\sf W}_{ij}$ all zero.
If the distribution of angles between the relative velocities of two consecutive collisions is (nearly) isotropic, the two average diagonal elements are (approximately) equal, so that
\begin{align}
{\sf W}_{ij} = 
\bar{w}{\bf 1} \delta_{ij}
\label{eq:w0}
~,
\end{align}
where $\delta_{ij}$ is the Kronecker delta.
The initial distribution used in the iteration process is a product of Dirac delta functions at the average value $\bar{w}$ for the diagonal elements and at zero for the off-diagonal elements.
In a similar way, an exponential distribution function can be used, with average $\bar{w}$, to test the sensitivity to the width of the distribution.

Using Eqs.~(\ref{eq:mathcalWa}) and (\ref{eq:mathcalWt}), one finds that after the collision and free flight,
 in the basis consisting of $\hat{\vec{v}}_{ij}$ and the $d-1$ vectors orthogonal to it, the values of ${\sf W}_{kl}$ have changed according to
\begin{widetext}
\begin{align}
{\sf W}_{kl}' = 
\begin{cases}
\left(
\begin{array}{cc}
(\bar{w}+\bar\nu\tau_k)&0\\
0& (\frac12 \bar{w} +\bar\nu\tau_k) {\bf 1}_{d-1}
\end{array}
\right) & 
\ifff k=l=i 
\vee k=l=j~,
\\[4ex]
\left(
\begin{array}{cc}
0&0\\
0& -\frac12 \bar{w}  {\bf 1}_{d-1}
\end{array}
\right) &
\ifff (k,l)=(i,j)
\vee (k,l)=(j,i)~,
\\[4ex]
\phantom{,}\bar{w}\,{\bf 1}\,\delta_{kl} &
\ifff k\neq i,j 
\vee l\neq i,j~,
\end{cases}
\label{eq:barw}
\end{align}
\end{widetext}
where ${\bf 1}_{d-1}$ denotes the $(d-1)$-dimensional identity matrix.
This equation implies a distribution for the elements of ${\sf W}'_{ij}$ expressed in the basis belonging to the next collision, which consists of $\hat{\vec{v}}_{ij}'$ and the $d-1$ vectors orthogonal to it, $\hat{\vec{v}}'_{ij\perp}$.
The new distribution of the matrix elements is the distribution of {${\sf W}_{kl}$} in the coordinates of the next collision, {${\mathcal R}_i\cdot {\sf W}'_{kl}\cdot{\mathcal R}_i^T$}, where {${\mathcal R}_i$} is the rotation matrix associated with the rotation from the coordinate system using the post-collisional relative velocity of a collision to the system using the pre-collisional relative velocity of the next collision of the same particle.
In two dimensions, this matrix is characterized by the angle $\phi_i$ between the relative velocities at the two collisions,
\begin{align}
{\mathcal R}_i = 
\left(\begin{array}{cc}
\cos\phi_i & \sin\phi_i\\
-\sin\phi_i & \cos\phi_i
\end{array}\right)~.
\end{align}
The distribution of this angle depends on the velocity of the particle between the two collisions.
In three dimensions, the angle $\phi_i$ is to be replaced by two angles.

From this approximation a distribution function of $w_{\perp\perp}$ can be found, which depends on $\bar{w}$.
At the next collision, it is, for $d=2$, equal to the distribution of
\begin{align}
w_{\perp\perp}'& =  \frac12 \hat{\vec{v}}'_{ij}\cdot ({\sf W}'_i + {\sf W}'_j ) \cdot \hat{\vec{v}}'_{ij}\\
& =  \frac12 (0,1) \cdot ({\mathcal R}_i \cdot {\sf W}'_i \cdot {\mathcal R}_i^T
+ {\mathcal R}_j \cdot {\sf W}'_j \cdot {\mathcal R}_j^T) \cdot (0,1)\\
& =  \bar{w} \left[1- \frac14 (\cos^2\phi_i+ \cos^2\phi_j)\right] + \bar\nu\tau_+~.
\label{eq:reswperpperp}
\end{align}
The distribution of $w_{\perp\perp}$ can be approximated by the distribution of the right-hand side of the equation.
For $d=3$, the rotation matrix is more complicated, and so Eq.~(\ref{eq:reswperpperp}) becomes more complicated.
The resulting expressions are not reproduced here.

From Eq.~(\ref{eq:barw}) one can find the difference between the average traces of ${\mathcal W}'$ and ${\mathcal W}$.
One finds,
\begin{align}
\langle{\trace({\mathcal W}')}\rangle - \langle \trace({\mathcal W}) \rangle = 2 d - (d -1) \bar{w}~,
\end{align}
where the notation $\langle . \rangle$ denotes the ensemble average.
As the average trace must not change, one finds for $\bar{w}$ the approximation
\begin{align}
\bar{w}^{(0)} = \frac{2 d}{d-1}=
\begin{cases}
4& \iff d=2~,\\
3& \iff d=3~.
\end{cases}
\label{eq:barw0}
\end{align}
Note that this result for $\bar{w}$ would be the same if an initial exponential distribution were used for the diagonal elements.
The resulting distribution function for $w_{\perp\perp}$, however is different in that case.
The distribution function implied by Eq.~(\ref{eq:reswperpperp}) for $w_{\perp\perp}$ at the next collision can be used to estimate the Kolmogorov-Sinai entropy from Eqs.~(\ref{eq:gemiddeld}) and (\ref{eq:sf}).
The involvement of these expressions was already anticipated in \cite{logtermen} by Dorfman, who predicted extra contributions to $B$ of $\ln 4$ and $\ln 3$ based on summations.

The approximation so far is fairly crude.
The nonlinearity of Eq.~(\ref{eq:mathcalWa}) has been partially neglected by using the averages of the off-diagonal elements.
In the second term in the calculation of the trace in Eq.~(\ref{eq:trace}) only the block diagonal terms, those of the form $\hat{\vec{v}}_{ij \perp}\cdot {\sf W}_{i}^2 \cdot \hat{\vec{v}}_{ij \perp}$, are involved.
In reality, since ${\mathcal W}$ is symmetric, the terms involving off-diagonal elements will also produce negative contributions to the average of the trace in Eq.~(\ref{eq:trace}).

A better approximation of the average value can be found by iterating the equation for the distribution a second time.
The distribution of ${\mathcal R}_i\cdot {\sf W}'_{kl}{\mathcal R}_i^T$ can be used to calculate the trace of ${\mathcal W}''$.
The colliding particles are uncorrelated before the collision, but not independent of the particles they encountered before.
These particles, which are not directly involved in the collision, now contribute to the change in the trace, through the second term on the right-hand side of Eq.~(\ref{eq:trace}).
 I find that for $d=2$ the trace in the low density approximation satisfies
\begin{widetext}
\begin{align}
\lefteqn{
\langle\trace({\mathcal W}'')\rangle-\langle\trace({\mathcal W}')\rangle =  - 3 \bar{w}
 + \null}& \nonumber\\
&
\left\langle\frac{\left[2 \bar\nu^2 (\tau_i^2 + 4 \tau_i\tau_j + \tau_j^2) + 10 \bar\nu(\tau_i + \tau_j) \bar{w}
+ 8 \bar{w}^2 - 2 \bar{w} \bar\nu( \tau_i \cos^2 \phi_j + \tau_j\cos^2\phi_i)
- \bar{w}^2 (\cos^2 \phi_i + \cos^2 \phi_j)\right]
}{
\left[{2\bar\nu (\tau_i+\tau_j) + 4 \bar{w} - \cos^2\phi_i - \cos^2\phi_j}\right]
}
\right\rangle
~.
\label{eq:dtrace}
\end{align}
\end{widetext}
This yields a result for $\bar{w}$ that is significantly different from Eq.~(\ref{eq:barw0}).

More iterations would produce more terms and will further reduce the value found for $\bar{w}$, converging to the exact result.
A similar but far more complicated expression can be found for $d=3$ from Eqs.~(\ref{eq:trace}), (\ref{eq:barw}), and the general form of the three-dimensional rotation matrix.
The results would improve if the distribution were iterated repeatedly, but this would produce expressions of complexity increasing exponentially with the number of iterations.
One more iteration would add four angles and four free-flight times to the expression in Eq.~(\ref{eq:dtrace}).
After the second iteration, the expression for $w''_{\perp\perp}$ is quite complicated and contains twelve correlated variables, six rotation angles and six free flights.
Also, the second iteration already produces a reasonable result.
I therefore continue using the distribution of $w'_{\perp\perp}$, but with the value of $\bar{w}$ found from Eq.~(\ref{eq:dtrace}).

The integrations over the distributions of $\phi_i, \phi_j, \tau_i$, and $\tau_j$ can be done numerically.
The change in the trace is zero for for $\bar{w}$ equal to
\begin{align}
\bar{w}^{(1)}_1  =
\begin{cases}
3.009  & \iff d=2~,\\
2.107  & \iff d=3~.
\end{cases}
\label{eq:resbarw}
\end{align}
The subscript index is introduced to indicate the weight given to the off-diagonal terms.

\subsubsection{Off-diagonal elements from earlier collisions}

If the contributions to the trace from the off-diagonal elements involving the other particles from the previous collisions, through the second term on the right-hand side of Eq.~(\ref{eq:trace}), are ignored, the result is changed significantly.
In this case,
\begin{align}
\bar{w}^{(1)}_0 = 
\begin{cases}
3.408 & \iff d=2~,\\
2.639 & \iff d=3~.
\end{cases}
\end{align}
At a collision between $i$ and $j$, the off-diagonal elements between particles $i$ and $k$ produce significant changes to the diagonal elements of ${\sf W}'_{kk}$.
It is therefore expected that contributions from particles involved in collisions before the previous collision will also be significant.
Also, if the other particle from the previous collision of particle $i$ has collided since, this has an effect on ${\sf W}_i$.

At a collision between particles $q$ and $r$, the change in the trace of ${\mathcal W}$, calculated in Eq.~(\ref{eq:trace}), is affected by the elements of ${\mathcal W}$ between $q$ and $r$, and other particles.
After a collision between two particles, nonzero off-diagonal elements exist between these particles.
After a collision between $i$ and $j$,
off-diagonal elements between particles $i$ and $k$ generate off-diagonal elements between $j$ and $k$, due to the exchange between the $\delta\vec{v}_i$ and $\delta\vec{v}_j$.
If nonzero off-diagonal elements exist between $i$ and $k$ as well as $j$ and $l$ before the collision, after the collision nonzero elements will exist between $k$ and $l$.
A diagrammatic representation of the collision sequence is shown in Fig.~\ref{fig:history}.

\begin{figure}[t]
\begin{center}
\includegraphics[width=6.0cm]{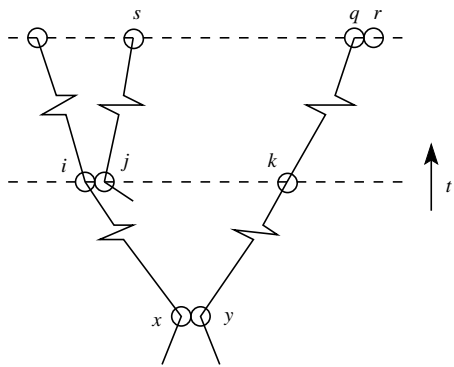}
\end{center}
\caption{\label{fig:history}
A diagrammatic representation of a series of collisions.
At the collision between particles $q$ and $r$, the off-diagonal elements between $q$ and particle $s$ contribute to the change in the trace.
These elements date from the collision between particles $x$ and $y$, in the common history of $q$ and $s$.
At some point in the past, there were elements between a particle $i$ in the history of $s$ and a particle $k$ in the history of $q$ when $j$, another particle in the history of $s$, collided with $i$.
After the collision there were elements between $j$ and $k$, which, through more collisions, eventually lead to elements between $s$ and $q$.
The size of the elements between $s$ and $q$ can be estimated using Eqs.~(\ref{eq:nietswitch}) and (\ref{eq:welswitch}).
}
\end{figure}

In order to estimate how much such terms contribute to the change in the trace at a collision involving particle $k$, the typical magnitude of the off-diagonal elements at a collision must be investigated.
One may estimate the typical changes in the off-diagonal blocks ${\sf W}_{ik}$ and ${\sf W}_{jk}$ at a collision between $i$ and $j$, by estimating the changes in the trace of the off-diagonal blocks.
The typical size of the off-diagonal elements can be characterized by the trace of the off-diagonal block ${\sf W}_{ik}$.
The diagonal elements of the off-diagonal blocks can be found from Eq.~(\ref{eq:offdiagel}).
Using the fact that ${\sf W}_{ij}$ before the collision is zero, one finds, in two dimensions, for the elements of ${\mathcal W}$ between $\delta\vec{r}_i$ and $\delta\vec{v}_k$
\begin{widetext}
\begin{align}
\label{eq:offdiag1}
\hat{\vec{v}}_{ij}\cdot {\sf W}'_{ik}\cdot \hat{\vec{e}} &= \hat{\vec{v}}_{ij}\cdot {\sf W}_{ik}\cdot \hat{\vec{e}}
- \frac{(\hat{\vec{v}}_{ij}\cdot {\sf W}_{i} \cdot \hat{\vec{v}}_{ij \perp})[\hat{\vec{v}}_{ij \perp} \cdot ({\sf W}_{ik}-{\sf W}_{jk})\cdot \hat{\vec{e}}] }{\hat{\vec{v}}_{ij \perp}\cdot ({\sf W}_{i} + {\sf  W}_{j})\cdot \hat{\vec{v}}_{ij \perp}}~,\\
\hat{\vec{v}}_{ij\perp}\cdot {\sf W}'_{ik}\cdot \hat{\vec{e}} &= \frac12 \hat{\vec{v}}_{ij \perp}\cdot ({\sf W}_{ik} - {\sf W}_{jk})\cdot \hat{\vec{e}}
- \frac{[\hat{\vec{v}}_{ij \perp}\cdot ({\sf W}_{i} - {\sf W}_j) \cdot \hat{\vec{v}}_{ij \perp}][\hat{\vec{v}}_{ij \perp} \cdot ({\sf W}_{ik}-{\sf W}_{jk})\cdot \hat{\vec{e}}] }{2 \hat{\vec{v}}_{ij \perp}\cdot ({\sf W}_{i} + {\sf  W}_{j})\cdot \hat{\vec{v}}_{ij \perp}}~.
\label{eq:offdiag2}
\end{align}
\end{widetext}
Here, $\hat{\vec{e}}$ can be any vector in two dimensions.
If ${\sf W}_{ik}$ has nonzero elements, then ${\sf W}_{jk}$ does not, since the particles $i$ and $j$ were uncorrelated before the collision.
If ${\sf W}_{ik}$ has nonzero elements after the collision, both ${\sf W}'_{ik}$ and ${\sf W}'_{jk}$ have nonzero elements.

From Eq.~(\ref{eq:offdiag1}) the traces after the collision may be found,
\begin{widetext}
\begin{align}
\vphantom{\frac{()\sf W}{()\sf W}} \trace({\sf W}'_{ik}) &=
\hat{\vec{v}}_{ij}\cdot {\sf W}_{ik} \cdot \hat{\vec{v}}_{ij} 
- \frac{(\hat{\vec{v}}_{ij}\cdot {\sf W}_{i} \cdot \hat{\vec{v}}_{ij \perp}) ( \hat{\vec{v}}_{ij \perp}\cdot {\sf W}_{ik} \cdot \hat{\vec{v}}_{ij}) } {\hat{\vec{v}}_{ij \perp}\cdot ({\sf W}_{i} + {\sf  W}_{j})\cdot \hat{\vec{v}}_{ij \perp}}\label{eq:wik1}
\nonumber\\
&
\vphantom{\frac{\sf W}{\sf W}}\phantom{=}\null+
\frac12 (\hat{\vec{v}}_{ij \perp}\cdot {\sf W}_{ik} \cdot \hat{\vec{v}}_{ij \perp})\left[ 1 - \frac{\hat{\vec{v}}_{ij \perp}\cdot ( {\sf W}_{i} - {\sf  W}_{j})\cdot \hat{\vec{v}}_{ij \perp}} {\hat{\vec{v}}_{ij \perp}\cdot ({\sf W}_{i} + {\sf  W}_{j})\cdot \hat{\vec{v}}_{ij \perp}}\right]
~,\\[2ex]
\vphantom{\frac{()\sf W}{()\sf W}} \trace({\sf W}'_{jk}) &=
\null
- \frac{(\hat{\vec{v}}_{ij}\cdot {\sf W}_j \cdot \hat{\vec{v}}_{ij \perp})( \hat{\vec{v}}_{ij \perp}  \cdot {\sf W}_{ik} \cdot \hat{\vec{v}}_{ij})  }
{\hat{\vec{v}}_{ij \perp}\cdot ({\sf W}_{i} + {\sf  W}_{j})\cdot \hat{\vec{v}}_{ij \perp}}
- \frac12 (\hat{\vec{v}}_{ij \perp}\cdot {\sf W}_{ik} \cdot \hat{\vec{v}}_{ij \perp})\left[ 1 - \frac{\hat{\vec{v}}_{ij \perp}\cdot ( {\sf W}_{i} - {\sf  W}_{j})\cdot \hat{\vec{v}}_{ij \perp}} {\hat{\vec{v}}_{ij \perp}\cdot ({\sf W}_{i} + {\sf  W}_{j})\cdot \hat{\vec{v}}_{ij \perp}}\right]
~.\label{eq:wik2}
\end{align}
\end{widetext}
It is fair to assume that the off-diagonal elements of ${\sf W}_{ik}$ tend to be smaller than the diagonal elements.
Also, the diagonal elements of ${\sf W}_i - {\sf W}_j$ are, typically, much smaller than the diagonal elements of ${\sf W}_i + {\sf W}_j$.
The terms with a quotient of these can therefore be neglected in this rough estimation.
Further, the $d$ diagonal elements of ${\sf W}_{ik}$ are typically of the same size.
These approximations leave us with the general expressions
\begin{align}
\label{eq:nietswitch}
\trace({\sf W}'_{ik}) &\approx \frac{d+1}{2d} \trace({\sf W}_{ik})~,\\[1ex]
\trace({\sf W}'_{jk}) &\approx \frac{d-1}{2d} \trace({\sf W}_{ik})~.
\label{eq:welswitch}
\end{align}

In addition, if both ${\sf W}_{ik}$ and ${\sf W}_{jl}$ have nonzero elements, ${\sf W}'_{kl}$ also has nonzero elements, which are due to the second term on the right-hand side of Eq.~(\ref{eq:offdiagel}).
The off-diagonal elements generated in this way are small compared to the elements generated from Eqs.~(\ref{eq:nietswitch}) and (\ref{eq:welswitch}).
In fact, they are smaller than the terms neglected from Eqs.~(\ref{eq:wik1}) and (\ref{eq:wik2}), because they contain products of the off-diagonal elements which are small compared to the diagonal elements.
As they appear quadratically in the change in the trace [see Eq.~(\ref{eq:trace})], they may be neglected, despite their quadratically larger number.

From Eqs.~(\ref{eq:nietswitch}) and (\ref{eq:welswitch}) an estimate can be made of the contributions to the change in the trace in Eq.~(\ref{eq:trace}) from collisions before the previous collision, compared to the contributions from just the previous collisions.
The ratio between the total contributions from off-diagonal elements to the change in the trace and the contributions from just the previous collisions is denoted by $\alpha$.

In every collision in the history, off-diagonal elements are created between the two colliding particles and existing elements are reduced in magnitude and passed on according to Eqs.~(\ref{eq:nietswitch}) and (\ref{eq:welswitch}).
In order to estimate the consequences of ${\sf W}_{qs}$ at a collision between $q$ and $r$ (see Fig..~\ref{fig:history}), one has to find the path through which information was passed on from the collision between particles $x$ and $y$ to the present collision between $q$ and $r$ as well as the path from the collision between $x$ and $y$ to the particle $s$ at the time of the collision between $q$ and $r$, following a sequence of collisions, through which the off-diagonal element between particles $q$ and $s$ is affected.
To this path belongs an approximate reduction of the size of the off-diagonal element, a product of factors of $(d+1)/(2d)$ or $(d-1)/(2d)$ for each collision in the paths.
If the path continues with the same particle, there is a factor of $(d+1)/(2d)$.
If it switches to the other particle, the factor is $(d-1)/(2d)$.

Every different product with the same number of factors follows a different path of that length, and hence belongs to a different present particle.
The product of the two factors of two paths starting from $x$ and $y$ gives the order of magnitude of the off-diagonal element between $i$ and $k$ particles.
The square of this factor then gives the relative size of the contribution to the trace at the collision between $i$ and $j$.
If a collision between two particles is now $p$ collisions ago, then, on average, the other part of the future of that collision has also had $p$ collisions.
Summing over all the different paths of length $p$, one finds that the relative contribution from collisions that occurred $p$ collisions before the previous collision can be approximated as
\begin{align}
\alpha_p \approx \left( \frac{d^2 + 1}{2d^2} \right)^{2p}~.
\end{align}
Summing over all $p$ gives the estimate
\begin{align}
\alpha  = \sum_p \alpha_p \approx \frac{4d^4}{(3d^2+1) (d^2-1)}~.
\end{align}

The contributions from the previous collisions in Eqs.~(\ref{eq:trace}) and (\ref{eq:dtrace}) can be multiplied by $\alpha$, to find an estimate for the total contribution of all particles with which $i$ and $j$ have a common history.
This is an admittedly crude estimate, yet should give better results than just neglecting the history before the previous collision.
The terms in Eq.~(\ref{eq:dtrace}) that are due to the off-diagonal block between the colliding particles and other particles may be multiplied by $\alpha$.
With this correction it is found that
\begin{align}
\bar{w}_\alpha^{(1)} =
\begin{cases}
2.929  & \iff d=2~,\\
1.947  & \iff d=3~.
\end{cases}
\label{eq:estbarw}
\end{align}

The distribution function of $w_{\perp\perp}$ contains an uncertainty in its width, which affects the results of the calculation.
When starting from the average, with every next iteration of the equation for the distribution function, the distribution becomes wider.
$w_{\perp\perp}$ looks like a sum of several weighted free-flight times for each particle.
If one starts from exponentially distributed diagonal elements, rather than simply the averages, the distribution becomes narrower with every iteration.
By starting from an exponential distribution, one may estimate the consequences of the width of the distribution of the elements.
With an initial exponential distribution, one finds 
\begin{align}
\bar{w}_\alpha^{(1)e} =
\begin{cases}
2.426  & \iff d=2~,\\
1.676  & \iff d=3~.
\end{cases}
\label{eq:estbarwexp}
\end{align}

By substituting the distribution function induced by one iteration, together with the average, into Eq.~(\ref{eq:gemiddeld}) one can now estimate the Kolmogorov-Sinai entropy.

\section{\label{sec:results}Results and discussion}

As the free-flight times are inversely proportional to the density, $w_{\perp\perp} \bar\tau^{d-1}$ will be inversely proportional to $n^{d-1}$.
This leads to a general form for the Kolmogorov-Sinai entropy,
\begin{align}
h_{\mathrm{KS}} = N \bar\nu A [-\ln (n a^d) + B]~.
\end{align}
In earlier calculations \cite{lagedichtheid}, $A$ was calculated accurately.
The results for $B$, however, are unsatisfactory.
The values of $A$ are easily found from Eqs.~(\ref{eq:gemiddeld}), (\ref{eq:sf}), and the dependence of the collision frequency on $n$,
\begin{align}
A = \frac{d-1}{2}
\label{eq:Aconstante}
~.
\end{align}
If $w_{\perp\perp}$ is taken equal to $\tau_+$, the results for $B$ of \cite{lagedichtheid} are reproduced,
\begin{align}
\tilde{h}_{\mathrm{KS}} = \frac{N\bar\nu}{2} \left\langle \ln \left[\left(\frac{2 v_{ij} \tau_+}{a}\right)^{d-1} \cos^{d-3}\theta\right]\right\rangle~.
\label{eq:henkenbob}
\end{align}
This yields
\begin{align}
\tilde{B}  \approx 
\begin{cases}
\phantom{-}0.209  & \iff d=2~,\\
-0.583 & \iff d=3~.
\end{cases}
\label{eq:Bconstantelagedichtheid}
\end{align}

From molecular-dynamics simulations the Kolmogorov-Sinai entropy has been calculated \cite{lagedichtheid,christinapriv}.
It is found that
\begin{widetext}
\begin{align}
h_{\mathrm{KS}}^{\mathrm s} =
\begin{cases}
(0.499 \pm 0.001)  N\bar\nu \left(-\ln n a^d + 1.366 \pm 0.005\right) & \ifff d=2~,\\
(1.02 \pm 0.02) N\bar\nu \left( -\ln n a^d + 0.29 \pm 0.01 \right) & \ifff d=3~.
\end{cases}
\label{eq:ksentropiemds}
\end{align}
\end{widetext}

In the calculation presented here, Eq.~(\ref{eq:henkenbob}) has to be amended, to become
\begin{align}
h_{\mathrm{KS}} = \frac{N\bar\nu}{2} \left\langle \ln \left[w_{\perp\perp}\left(\frac{2 v_{ij}\bar\tau}{a}\right)^{d-1} \cos^{d-3}\theta\right]\right\rangle~.
\end{align}
From Eqs.~(\ref{eq:reswperpperp}) and (\ref{eq:resbarw}), one finds, after numerical integration, that
\begin{align}
B^{(1)}_1 \approx 
\begin{cases}
1.592 & \iff d=2~,\\
0.476 & \iff d=3~.\\
\end{cases}
\label{eq:B}
\end{align}

If the contributions from the off-diagonal elements in Eq.~(\ref{eq:trace}) are increased by the estimate of the remaining terms, through a factor of $\alpha$, the results change to
\begin{align}
B^{(1)}_\alpha \approx
\begin{cases}
1.572 & \iff d=2~,\\
0.427 & \iff d=3~.
\end{cases}
\label{eq:estB}
\end{align}
This more closely reproduces the simulation results shown in Eq.~(\ref{eq:ksentropiemds}).

After every extra iteration in the calculation, the distribution becomes wider and therefore the average of the logarithm of $w_{\perp\perp}$ becomes smaller compared to the logarithm of the average.
Due to this, cutting off the process after two iterations produces a result for the Kolmogorov-Sinai entropy which is too high.
Note that also a wider spread of the off-diagonal elements leads to larger contributions from the off-diagonal terms in Eq.~(\ref{eq:trace}), and therefore to a smaller $\bar{w}$, which yields a smaller value for $B$.
Equation~(\ref{eq:B}) threfore gives an upper bound for $B$.

By starting from a wider distribution of diagonal elements instead of a product of Dirac delta functions, an estimate can be made of the effects of the width of the distribution.
From the results of an exponential initial distribution, Eq.~(\ref{eq:estbarwexp}), an estimation is found,
\begin{align}
B^{(1)e}_\alpha \approx
\begin{cases}
1.370 & \iff d=2~,\\
0.273 & \iff d=3~.\\
\end{cases}
\label{eq:estBexp}
\end{align}

From these two estimated bounds, a final estimate of $B$ may be made, including error bounds,
\begin{align}
B =
\begin{cases}
1.47 \pm 0.11 & \iff d=2~,\\
0.35 \pm 0.08 & \iff d=3~.\\
\end{cases}
\label{eq:finB}
\end{align}
The errors could be reduced by using distribution functions for $w_{\perp\perp}$ that have been iterated a larger number of times.
The values of the Kolmogorov-Sinai entropy found in the molecular dynamics simulations, Eq.~(\ref{eq:ksentropiemds}), are well within the error bounds of Eq.~(\ref{eq:finB}).

\section{Conclusions}

In this paper, an estimation using kinetic theory is presented of the Kolmogorov-Sinai entropy of dilute hard sphere gases in equilibrium.
Kinetic theory has been applied before to calculate chaotic properties \cite{3d,3dposch,lagedichtheid}, such as the Lyapunov spectrum of the high dimensional Lorentz gas \cite{onslorentz}.
In systems with escape, the Kolmogorov-Sinai entropy has also been connected to transport coefficients \cite{GN2,transport1,transport2,transport3}, but in the system investigated here, it is just equal to the sum of the positive Lyapunov exponents.
It is known that the leading orders are of the form $N \bar\nu A (-\ln n + B)$\cite{lagedichtheid}.

An nonlinear integral equation was derived for the joint distribution function of the elements of the inverse of the radius of curvature tensor.
This equation was approximately solved by the use of an iterative method.
$B$ was estimated from the solution in a satisfactory way, with results which are consistent with simulation results.
It was found that, $B = 1.47 \pm 0.11$ for $d=2$ and $B = 0.35 \pm 0.08$ for $d=3$.
The approximations made are systematic, and the results can be further improved by performing more iterations of the distribution.
The values for $B$ found in the present calculation are in good agreement with the results from molecular-dynamics simulations \cite{christinapriv}.
Also, an upper bound was found for $B$, that is, $B < 1.592$ for $d=2$, and $B < 0.476$ for $d=3$.

The smaller Lyapunov exponents of this system
are proportional to $\bar\nu$.
The ones which are not due to Goldstone modes contribute significantly to $B$ in the Kolmogorov-Sinai entropy.
The calculation of $B$ presented here shows effects which affect the behavior of these Lyapunov exponents to the leading order \cite{tbp}.

It should be noted that effects such as the ones described here do not affect the Lyapunov spectrum of the high-dimensional Lorentz gas, calculated in reference \cite{onslorentz}, because the scatterers in that system are uniformly convex.
However, they are generic for the Lyapunov spectra of systems consisting of many particles.

\begin{acknowledgments}
I would like to thank Henk van Beijeren for many helpful discussions and for making me explain things properly.

This work was supported by the {\em Collective and cooperative statistical physics phenomena} program of FOM (Fundamenteel Onderzoek der Materie).
\end{acknowledgments}


\end{document}